\documentclass[fleqn,usenatbib,useAMS]{mnras}
\usepackage[T1]{fontenc}
\usepackage{ae,aecompl}
\usepackage{graphicx}	
\usepackage{amsmath}	
\usepackage{amssymb}	
\usepackage{bm}		
\usepackage{epsfig}

\usepackage{txfonts} 

\PassOptionsToPackage{pdfpagelabels=false}{hyperref} 

\usepackage{color}

\newcommand{\camcs}{App. Math. Comp. Sci.}

\newcommand{\lcdm}{$\Lambda$CDM}

\newcommand{\OmL}{\Omega_{\mathrm{\Lambda}}}
\newcommand{\OmM}{\Omega_{\mathrm{M}}}
\newcommand{\OmMo}{\Omega_{\mathrm{M},0}}
\newcommand{\Ombo}{\Omega_{\mathrm{b},0}}

\newcommand{\OmLo}{\Omega_{\mathrm{\Lambda},0}}
\newcommand{\msol}{M_{\odot}}
\newcommand{\sig}{\sigma_{8}}
\newcommand{\rvir}{R_{\rmn{vir}}}

\newcommand{\EinRadMed}{\theta_{\mathrm{E,med}}}
\newcommand{\EinRadEff}{\theta_{\mathrm{E,eff}}}
\newcommand{\LMrel}{L$_{X}$--M}

\newcommand{\DMsim}{{\it DM}}
\newcommand{\NRsim}{{\it NR}}
\newcommand{\CSFsim}{{\it CSF}}
\newcommand{\AGNsim}{{\it AGN}}

\newcommand*\diff{\mathop{}\!\mathrm{d}}

\title[Marginal likelihood from cosmological simulations]{Simulation-based marginal likelihood for cluster strong lensing cosmology}

\author[M. Killedar et al.]{
M.~Killedar,$^{1,2,3}$\thanks{E-mail: madhura.killedar.astro@gmail.com (MK)}
S.~Borgani,$^{3,4,5}$ 
D.~Fabjan,$^{6,4}$ 
K.~Dolag,$^{2,7}$
G.~Granato,$^{3,4}$
\newauthor 
M.~Meneghetti,$^{8,9,10}$ 
S.~Planelles$^{3,4,5,11}$ \& 
C.~Ragone-Figueroa$^{4,12}$\\
$^{1}$Centre for Translational Data Science, The University of Sydney, NSW 2006, Australia\\
$^{2}$Universit\"ats-Sternwarte M\"unchen, Scheinerstrasse 1, D-81679, M\"unchen, Germany\\
$^{3}$Dipartimento di Fisica dell'Universit\`{a} di Trieste, Sezione di Astronomia, Via Tiepolo 11, I-34131 Trieste, Italy\\
$^{4}$INAF - Osservatorio Astronomico di Trieste, Via G.B. Tiepolo 11, I-34131 Trieste, Italy\\
$^{5}$INFN - National Institute for Nuclear Physics, Via Valerio 2, I-34127 Trieste, Italy\\
$^{6}$Faculty of Mathematics and Physics, University of Ljubljana, Jadranska 19, 1000 Ljubljana, Slovenia\\ 
$^{7}$Max-Planck-Institut f\"ur Astrophysik, Garching, Germany\\
$^{8}$INAF - Osservatorio Astronomico di Bologna, Via Ranzani 1, I-40127 Bologna, Italy\\
$^{9}$INFN - Sezione di Bologna, Viale Berti Pichat 6/2, I-40127 Bologna, Italy\\
$^{10}$Jet Propulsion Laboratory, 4800 Oak Grove Dr. Pasadena, CA 91109, USA\\
$^{11}$Departament d'Astronomia i Astrof\'{\i}sica, Universitat de Val\`encia, 46100- Burjassot (Valencia), Spain\\
$^{12}$Instituto de Astronom\'ia Te\'orica y Experimental (IATE), Consejo Nacional de Investigaciones Cient\'ificas y T\'ecnicas de la Rep\'ublica Argentina\\ (CONICET), Observatorio Astron\'omico, Universidad Nacional de C\'ordoba, Laprida 854, X5000BGR, C\'ordoba, Argentina
}

\date{Accepted 2017 August 31. Received 2017 August 31; in original form 2015 July 17}
\pubyear{2017}

\begin{document}
\label{firstpage}
\pagerange{\pageref{firstpage}--\pageref{lastpage}}
\maketitle


\begin{abstract}
Comparisons between observed and predicted strong lensing properties of galaxy clusters have been routinely used to claim either tension or consistency with \lcdm~cosmology. However, standard approaches to such cosmological tests are unable to quantify the preference for one cosmology over another. We advocate approximating the relevant Bayes factor using a marginal likelihood that is based on the following summary statistic: the posterior probability distribution function for the parameters of the scaling relation between Einstein radii and cluster mass, $\alpha$ and $\beta$. We demonstrate, for the first time, a method of estimating the marginal likelihood using the X-ray selected $z>0.5$ MACS clusters as a case in point and employing both N-body and hydrodynamic simulations of clusters. We investigate the uncertainty in this estimate and consequential ability to compare competing cosmologies, that arises from incomplete descriptions of baryonic processes, discrepancies in cluster selection criteria, redshift distribution, and dynamical state. The relation between triaxial cluster masses at various overdensities provide a promising alternative to the strong lensing test.
\end{abstract}


\section{Introduction}
The matter density parameter, $\OmM$, the vacuum density parameter, $\OmL$, the normalisation of the matter power spectrum, $\sig$, and the slope, $n$, of the power spectrum for the primordial density fluctuations ($P(k) \propto k^{n}$) have a strong influence on the redshift at which clusters form and the amount of time they are given to evolve until we observe them \citep[e.g.][for a review]{KravtsovBorgani12}. For clusters of a fixed mass at the time of observation, lower values of $\OmM$, $\OmL$ or $\sig$ or higher values of the spectral index, $n$, would require the host haloes to have formed earlier, and subsequently lead to higher concentrations \citep{CL96, TBW97, NFW97, W02, vdB02}. Galaxy clusters gravitationally lens and distort the images of background galaxies; their lensing efficiency is a powerful probe of cosmology with the ability to constrain the aforementioned structure formation parameters \citep{B98,TC01,B03,WBO04,Boldrin2016}. This is partly because the cosmological model determines the formation history of clusters, but also because the critical surface mass density for lensing, a function of the angular diameter distances between observer, lens and source(s), is also dependent on these cosmological parameters. However, cosmological distances play a secondary role compared to the mass distribution of clusters \citep{WH93, OTS01, HWY07}.  The earliest comparisons between simulated clusters and the observed frequency of arc-like lensed galaxy images in a cluster sample revealed an order of magnitude difference between the observations and \lcdm~predictions \citep{B98,L05}.
This discrepancy, dubbed the `arc-statistics problem', has the potential to be a point of tension for the standard \lcdm~model \citep[see][ for an overview]{Meneghetti13}. As such, many efforts have been made to provide explanations, beginning with discussions about the appropriate modelling of shape and redshift distribution of the background source galaxies \citep{BSW95,WBO04,HW05,L05,HWY07,Bayliss12}, complex structure in the lensing mass \citep[e.g.][]{Bayliss2014}, and the nature of dark matter \citep[e.g.][]{Mahdi2014}.
Cluster selection criteria are another complicating factor \citep{Sereno2015}. Strong-lensing selection will sample the high mass end of the cluster mass function \citep{CN07} and preferentially target clusters aligned along the major axes \citep{M93,Hennawi07,OB09,Men10b}. X-ray selection tends to create a sample including more merging clusters with complex morphology \citep{RT02,Planelles2009}, and yet high-concentration \citep{Rasia13} and a higher fraction of cool-core systems (Jones et al., in preparation).

Comparisons between simulated and observed clusters have been conducted at a range of cluster lens redshifts.  \citet{DHH04} found that arc statistics associated with low-redshift cluster lenses are consistent with observations, although \citet{Horesh11} maintain that observed number counts are higher than expected for clusters at $z\lesssim0.2$. All studies so far have found that the discrepancy remains at high redshift ($z\gtrsim0.6$) for the most massive clusters or is unclear due to small number statistics for high-mass simulated clusters \citep{DHH04,Horesh11,Men11}. We improve upon these works by increasing the simulated sample size somewhat. However, given the sensitivity of arc-statistics to the assumed properties of background sources and the lack of detailed observations required for complete mass profile models, we compromise by using Einstein radii as a proxy for cluster mass concentration.

The Massive Cluster Survey (MACS) is one of few samples that have a simple, well-defined selection function, high completeness and ample data for strong lensing analysis. There are also updates to some of the clusters' mass models care of the Cluster Lensing and Supernova survey with Hubble \citep[CLASH;][]{Postman2012,Zheng12,Coe13,Zitrin2015} and ongoing Frontier Fields programmes \footnote{\tt http://www.stsci.edu/hst/campaigns/frontier-fields/} \citep{Johnson2014}. The X-ray selected $z>0.5$ MACS cluster sample \citep{E07} has posed such a challenge due to large measured Einstein radii. However there are disagreements throughout the literature due the differing theoretical models and statistical methods. In the present work, we propose a Bayesian approach to the cosmological test using strong lensing properties of this sample, and clusters modelled within hydrodynamic simulations.

Concerns about the lack of baryonic heating and cooling mechanisms in early cosmological simulations led to a number of studies examining the effect of these processes on cluster density profiles and lensing efficiency \citep{Lewis00,P05,Rozo08,D10,Mead10,Cui12,Killedar12,Rasia13}. Together, they paint a complex picture of numerous counteracting baryonic effects. `Runaway' cooling flows in simulations have been found to steepen profiles and produce stronger lenses \citep{P05, Rozo08}. Yet, there is no significant link between ICM cooling signatures and lensing efficiency in strong lensing selected clusters \citep{Blanchard13}. This apparent discrepancy is resolved with the additional component of feedback mechanisms which temper the overproduction of stellar mass while simultaneously reducing the strong lensing efficiency \citep{Sijacki07,T11,McC10,Fabjan10,Mead10,Killedar12}. In \citet{Killedar12}, we found that the quantitative difference between cluster lensing properties in N-body and hydrodynamic simulations depends mildly on the redshift of the cluster lenses, and the results suggested that for relaxed clusters, the inclusion of baryons would not affect lensing efficiencies characterised by Einstein radii or tangential arcs. However, cluster selection criteria and characterisation of strong lensing efficiency affect the result of the comparison. Exploring in detail such effects is included in the analysis that we present here. We are not restricted to relaxed clusters in the present sample, although we investigate the consequences of dynamical selection. 
  
 The paper is structured as follows: in Section~\ref{Sec:Lensing} the basic theory and notation of gravitational lensing are introduced, along with the characterisation of the Einstein radius; in Section~\ref{Sec:MACS} the MACS high-z sample is described; the main details of the hydrodynamic simulations are briefly outlined in Section~\ref{Sec:HydroSims}; in Section~\ref{Sec:ML} we motivate a Bayesian approach and describe a method by which one estimates the marginal likelihood; in Section~\ref{Sec:Results} we demonstrate and discuss selection effects and uncertainties with regards to the modelling of baryonic processes; we discuss how our findings differ from previous strong lensing studies of the MACS clusters in Section~\ref{Sec:CompareLit}; in Section~\ref{Sec:triax} we consider the potential of triaxial mass models to provide an alternative scaling relation; we finally summarize our findings in Section~\ref{Sec:Conclude}.

Throughout the present work, the following values for cosmological parameters are adopted: present day vacuum density parameter, $\OmLo=0.76$; matter density parameter, $\OmMo=0.24$; baryon density parameter, $\Ombo=0.04$; Hubble constant $h=0.72$; normalisation of the matter power spectrum $\sig=0.8$; and primordial power spectrum $P(k) \propto k^{n}$ with $n=0.96$. Furthermore, characteristic overdensities at which cluster masses are provided, are assumed to refer to the critical cosmic density, $\rho_c=3H^2/(8\pi G)$.

\section{Strong Lensing Efficiency}\label{Sec:Lensing}

Throughout the present work, we refer to gravitational lensing quantities following the notation of \citet{SEF92}, and assuming the thin lens approximation. 
Note that from here onwards, the redshift of the background source galaxies is denoted $z_{s}$. Since the strong lensing properties of the observational sample are usually determined for galaxies at the fixed source redshift of $z_{s}=2$, we derive results for the same unique source redshift throughout this work.

\subsection{Gravitational lensing}
Images of a source at $z_{s}$ are highly magnified when they appear on a locus known as the {\it critical curve}, where the Jacobian of the lens mapping formally diverges.

The angular separation of highly-magnified and tangentially-sheared background galaxies has a
formal definition, which is strictly applicable only in the case of
axially symmetric lenses. This separation is defined by the Einstein radius:
\begin{equation}\label{Eqn:ERsymlens}
	\theta_{E}	= \sqrt{\frac{4GM}{c^{2}}\frac{D_{ds}}{D_{d}D_{s}}}\,,
\end{equation}
where $M$ denotes the enclosed mass and $D_{s}$, $D_{d}$ and $D_{ds}$ are the angular diameter distances from the observer to the source, from the observer to the lens, and from the lens to the source, respectively. However, galaxy clusters are not axially symmetric in general, and so critical curves are not circular. As such, the typical scale length may be characterised by the so-called `effective Einstein radius', $\EinRadEff$, according to:
\begin{equation}\label{eqn:effrad}
	 A=\pi \EinRadEff^{2}\,,
\end{equation}
where $A$ is the area enclosed within the tangential critical curve. This is the definition used, for example, by \citet{PH09}, \citet{Z11a} and \citet{Redlich12}, and is implemented throughout the present work (however, see Sec.~\ref{Sec:CompareLit}).

\section{The \texorpdfstring{$\bmath{\lowercase{z} > 0.5}$}{z>0.5} MACS sample}
\label{Sec:MACS}

MACS consists of the most X-ray luminous clusters, from which a 90 per cent complete sample of high-redshift clusters ($z>0.5$) were presented in \citet{E07}. Ultimately the clusters are chosen by following a flux-limit and redshift criterion, and so should not suffer from the lensing-selection bias. For the {\it ROSAT} All-Sky Survey Bright Source Catalogue (RASS BSC) the flux limit is $f_{X} > 1 \times 10^{-12}$ erg s$^{-1}$ cm$^{-2}$ in the 0.1--2.4 keV band. However, follow-up observations with {\it Chandra} found that the lowest flux cluster had $f_{X} = 0.8 \times 10^{-12}$ erg s$^{-1}$ cm$^{-2}$ in the same band; given the higher sensitivity of {\it Chandra}, we use this latter flux limit for our simulated cluster selection (see Sec.~\ref{SubSec:AGNLX}).
Masses within R$_{500}$ assuming sphericity and hydrostatic equilibrium were derived by \citet{Mantz2010}. Changes to the flux measurements from the 2009 January 21 {\it Chandra} calibration update mean that these masses are likely overestimated by 7--15 per cent, therefore throughout this work, we reduce the masses by 10 per cent as a rough guide, and estimate a 10 per cent uncertainty. Lensing data from the CLASH programme has led to new mass estimates for MACS J0717.5+3745, MACS J1149.5+2223, MACS J0744.83927, and MACS J0647.7+7015 \citep{Umetsu2016}.

The morphological codes listed in \citet{E07} suggest that half the clusters are unrelaxed (see further discussion in Sec.~\ref{SubSec:relax}). 
%
\begin{table*}
 \centering
   \caption{Properties of MACS $z>0.5$ cluster sample. Column 2: redshift. Column 3: Morphological code \citep[see][]{E07}. Columns  4--5: {\it Chandra} fluxes and luminosities in the 0.1--2.4 keV  band, quoted from \citet{E07}. Column 6: Masses within R$_{500}$  assuming sphericity and hydrostatic equilibrium are cited from \citet{Mantz2010}; 2009 January 21 {\it Chandra} calibration update fluxes mean that these are overestimated by 7--15 per cent. Column 7: Effective Einstein radius at $z_{s}=2$ (see text for details and  references) 
   }
  \label{Tab:MACSinfo}
 \begin{minipage}{120mm}
  \begin{tabular}{@{\extracolsep{\fill}}lccccccc@{\extracolsep{\fill}}}
  \hline
  MACS     		& z 	&	 Morph. 	&  f$_{X}$ 				& L$_{X}$ 		& M$_{500}$	& $\EinRadEff$	\\
       			& 	& 	& $10^{-12}$ erg s$^{-1}$cm$^{-2}$ & 10$^{44}$ erg s$^{-1}$& $10^{14}\msol$ 	& arcsec 	\\
  \hline
 J0018.5+1626 	& 0.5456 	& 3 		& $2.14\pm0.03$		& $19.6\pm0.3$		& $16.5\pm2.5$		& $24$ 	  \\
 J0025.4-1222  	& 0.5843 	& 3 		& $0.81\pm0.02$		& $8.8\pm0.2$		& $ 7.6\pm0.9$ 		& $30$	 \\
 J0257.1-2325 	& 0.5049 	& 2 		& $1.80\pm0.03$		& $13.7\pm0.3$		& $ 8.5\pm1.3$		& $39$	\\
 J0454.1-0300  	& 0.5377 	& 2 		& $1.88\pm0.04$		& $16.8\pm0.6$		& $11.5\pm1.5$		& $13$	\\
 J0647.7+7015	& 0.5907 	& 2 		& $1.49\pm0.03$		& $15.9\pm0.4$		& $10.9\pm1.6$		& $28$ 	  \\
 J0717.5+3745 	& 0.5458 	& 4 		& $2.74\pm0.03$		& $24.6\pm0.3$		& $15.0\pm2.85$ 		& $50.1$	 \\
 J0744.8+3927	& 0.6976 	& 2 		& $1.44\pm0.03$		& $22.9\pm0.6$		& $11.9\pm2.8$ 		& $23.3$	\\
 J0911.2+1746 	& 0.5049 	& 4 		& $1.00\pm0.02$		& $7.8\pm0.3$		& $ 9.0\pm1.2$ 		& $11$	\\
 J1149.5+2223 	& 0.5444 	& 4 		& $1.95\pm0.04$		& $17.6\pm0.4$		& $14.6\pm3.1$ 		& $21.5$ 	 \\
 J1423.8+2404 	& 0.5431 	& 1 		& $1.80\pm0.06$		& $16.5\pm0.7$		& $ 6.6\pm0.9$ 		& $17.8$ 	\\
 J2129.4-0741		& 0.5889 	& 3 		& $1.45\pm0.03$		& $15.7\pm0.4$		& $10.6\pm1.4$ 		& $21.8$	\\
 J2214.9-1359  	& 0.5027 	& 2 		& $1.85\pm0.03$		& $14.1\pm0.3$		& $13.2\pm2.3$ 		& $23$	\\
\hline
\end{tabular}
\end{minipage}
\end{table*}

The effective Einstein radii were originally presented in \citet{Z11a}. However a number of these have been revised with new imaging and spectroscopic data, as well as mass reconstruction techniques.
MACS J0647.7+7015 has been re-analysed by \citet{Coe13} as well as MACS J2129.4-0741, MACS J0744.83927 and MACS J1423.8+2404 by \citep{Zitrin2015} using HST data collected as part of the CLASH program. 
New models for MACS J1149.5+2223 and MACS J0717.5+3745 have been presented by \citet{Johnson2014} using data collected from the Frontier Fields programme. The Einstein radii at $z_{s}=2$ are found to be 21.5 and 50.1 arcsec respectively.
In the case of the `baby bullet' cluster MACS J0025.4-1222, the secondary critical curve was included in the analysis, despite being slightly separated from the primary critical curve. However in any case, simulated clusters undergoing a merger will show a large variation in strong lensing measurements depending on the projection; secondary critical curves will often be connected to the primary.
In the absence of more complete information, we estimate a 1 percent Gaussian uncertainty on the Einstein radii of MACS J1149.5+2223 and J0717.5+3745 \citep{Johnson2014}, and 10 percent uncertainty on all the other MACS high-z clusters (A. Zitrin, private communication). All relevant properties of the high-z MACS clusters are listed in Table~\ref{Tab:MACSinfo}.


\section{Cosmological Simulations}
\label{Sec:HydroSims}
The simulations analyzed here are the same as described in \citet{RagoneFigueroa13} and \citet{Planelles14}. In the following, we provide a short overview, while we refer to the above papers for a comprehensive description. 

\subsection{The set of simulated clusters}
Simulations have been carried out using the TreePM--SPH {\small GADGET--3} code, a newer version of the original {\small GADGET--2} code by \cite{S05} that adopted a more efficient domain decomposition to improve the work-load balance. A flat \lcdm~model whose cosmological parameters were chosen as follows: present day vacuum density parameter, $\OmLo=0.76$; matter density parameter, $\OmMo=0.24$; baryon density parameter, $\Ombo=0.04$; Hubble constant $h=0.72$; normalisation of the matter power spectrum $\sig=0.8$; and primordial power spectrum $P(k) \propto k^{n}$ with $n=0.96$.

Starting from a low-resolution cosmological box having size of 1~h$^{-1}$ Gpc, we selected 24 Lagrangian regions surrounding the most
massive clusters identified at $z=0$, all having virial mass of at least $10^{15}$h$^{-1}\msol$, plus further 5 Lagrangian regions surrounding clusters in the mass range (1--5)$\times 10^{14}$h$^{-1}\msol$ (see \citealt{Bonafede11} for details). Initial conditions are then generated by increasing mass resolution, and correspondingly adding higher frequency modes to the density fluctuation field, within these regions. Resolution is progressively degraded outside these regions, so as to save computational time while still providing a correct description of the large-scale tidal field. The Lagrangian regions were large enough to ensure that only high-resolution particles are present within five virial-radii of the central cluster.

Each Lagrangian region has been simulated in four different flavours: including only dark matter particles (\DMsim); with non-radiative hydrodynamics (\NRsim); including cooling star formation and supernova (SN) feedback (\CSFsim); and further including AGN feedback (\AGNsim). 

The basic characteristics of these re-simulation sets are described here below.
\begin{description}
\item[{\it DM}]: simulations including only dark matter particles, that in the high-resolution region have a mass  $m_{DM}=10^{9}h^{-1}\msol$. The Plummer--equivalent co-moving softening length for gravitational force in the high-resolution  region is fixed to $\epsilon_{Pl}=5 h^{-1}$ kpc physical at $z<2$  while being fixed to $\epsilon_{Pl} = 15   h^{-1}$ kpc comoving at higher redshift.\\
\item[{\it NR}]: non-radiative hydrodynamical simulations. Initial conditions for these hydrodynamical simulations are generated   starting from those of the DM-only simulations, and splitting each  particles in the high resolution region into one dark matter and one   gas particle, with their masses chosen so as to reproduce the assumed cosmic baryon fraction.  The mass of each DM particle is then
  m$_{\rm{DM}} = 8.47 \cdot 10^8 \, \rm{h}^{-1} \msol$ and the mass of  each gas particle is m$_{\rm{gas}} = 1.53 \cdot 10^8 \, \rm{h}^{-1}  \msol$. For the computation of the hydrodynamical forces we assume  the minimum value attainable by the SPH smoothing length of the  B-spline interpolating kernel to be half of the corresponding value  of the gravitational softening length.  No radiative cooling is included.\\
\item[{\it CSF}]: hydrodynamical simulations including the effect of
  cooling, star formation, chemical enrichment and SN feedback. Star formation is described
  through the effective model by \cite{SH03}. The effect of SN
  feedback is included by using galactic wind having a velocity of 500
  km s$^{-1}$. Chemical enrichment is described as in
  \cite{Tornatore2007} and includes the contributions from Type-Ia and
  Type-II SN, and of AGB stars.
\item[{\it AGN}]: the same as {\it CSF}, but with the additional
  effect of AGN feedback.  In the model for AGN feedback, released
  energy results from gas accretion onto super-massive black holes
  (SMBHs), that are initially seeded within resolved DM halos and
  later grow by gas accretion and merging with other BHs. The
  description of BH accretion and AGN feedback used in our simulations
  is largely inspired by that originally presented by \cite{SdMH05},
  with a number of modifications, whose details and motivation are
  explained in \citet{RagoneFigueroa13} (see also
  \citealt{Planelles14}).
\end{description}

\subsection{Properties of the simulated clusters}\label{SubSec:simclus}
The clusters within the simulated regions are identified as follows. Firstly, a standard {\it Friends-of-Friends} (FoF) algorithm is run over the dark matter particles in the high-resolution regions, using a linking length of 0.16 in units of the mean inter-particle separation. Within each FoF group, we identify the position of the particle with the minimum gravitational potential, which is then taken as the centre from where clusters are then identified according to a spherical overdensity (SO) method.
The mass, $M_{500}$, of each cluster is defined as the mass enclosed within the radius, $R_{500}$ at which the average density is 500 times the critical overdensity.

Throughout Sec.~\ref{Sec:Results}, we provide the results of analyses of simulated clusters chosen either by mass or X-ray luminosity within the 0.1--2.4 keV energy band. Since we aim to perform a self-consistent cluster selection with respect to the high-z MACS cluster selection criteria, we estimate the X-ray luminosity for simulated clusters within the \AGNsim~simulations, which are those producing a relation between X-ray luminosity and mass consistent with observational results \citep[][]{Planelles14}. The X-ray luminosity is computed by summing the contributions to the emissivity, $\epsilon_i$, carried by all the gas particles within $R_{500}$:
\begin{equation}
L_X=\sum_i\epsilon_i=\sum_i\,n_{{\rm e},i}n_{{\rm H},i}\Lambda(T_i,Z_i){\rm d}V_i\,,
\label{eq:emiss}
\end{equation}
where $n_{{\rm e},i}$ and $n_{{\rm H},i}$ are the number densities of electrons and of hydrogen atoms, respectively, associated with the $i$-th gas element of given density $\rho_i$, temperature $T_i$, mass
$m_i$, metallicity $Z_i$, and volume ${\rm d}V_i=m_i/\rho_i$. Furthermore, $\Lambda(T,Z)$ is the temperature- and metallicity-dependent cooling function computed within the $[0.1-2.4]$ keV energy band.

\subsection{Measuring Einstein radii}\label{SubSec:ERcalc}
The lensing mass includes all matter within two virial radii\footnote{The virial radius is defined as the smallest radius of a sphere centred on the cluster, for which the mean density falls below the virial overdensity. The virial overdensity is measured relative to the critical density and calculated using the fitting formula of \citet{BN98}.} of the cluster centre. Using the Fourier techniques outlined in \citet{Killedar12}, we determine the positions of critical points for a projected lens. Firstly, the projection is centered on the peak in the two-dimensional surface map, which is likely to reside in the largest critical curve within this field. Tangential critical points are identified within a square field of view (comoving 1.5 h$^{-1}$ Mpc across) on a fine 2048-pixel grid (giving an angular resolution of 0.1 arcsecs at $z=0.5$). As large substructures can also be present, we remove critical points associated with any distinct secondary critical curves, before measuring the Einstein radius. The effective Einstein radius is defined by equation~(\ref{eqn:effrad}) where $A$ is the angular area enclosed by the polygon bounded by the remaining critical points. 

Snapshots of the cosmological simulations are taken at fixed redshifts ($z=0.5$ and $z=0.6$) from which we may select galaxy cluster-scale objects as a representative description of lenses as predicted by \lcdm. However, the MACS cluster high-z sample span a range of redshifts. Thus quantities that describe strong lensing, which ultimately reflect the mass distribution in the inner regions of the lens, should be scaled in a way that makes the quantities sensitive to that mass, but robust to offsets in redshift between the lenses being compared. Rearranging equation~(\ref{Eqn:ERsymlens}) we find:
\begin{equation} \label{einradscale}
	\theta_{E} \sqrt{ \frac{D_{d}}{D_{ds}} } \propto \sqrt{M},
\end{equation}
which provides a rough scaling for a strong lensing quantity that scales with enclosed mass.

\section{Estimating the Marginal Likelihood} 
\label{Sec:ML}
Strong lensing efficiencies, as characterised by the Einstein radii, scale well with the mass of clusters at large overdensities \citep[see][]{Killedar12}.  If the $z>0.5$ MACS sample are, in fact, stronger lenses than predicted by the \lcdm~model, they will have larger Einstein radii for a given total mass at low overdensities (or a proxy thereof).

A Bayesian approach is advocated \citep[see e.g. ][]{Sivia1996, Trotta2008, Jenkins2011}, in which one determines the relative preference of two hypothetical cosmological models, $C_{1}$ and $C_{2}$, in light of the data $D$:
\begin{equation}
\label{eqn:cosmoprefer}
	\frac{P(C_{1}| D)}{P(C_{2}| D)} = R  \frac{P(C_{1})}{P(C_{2})}\,,
\end{equation}
where $P(C_{1})/P(C_{2})$ denotes the prior preference for $C_{1}$ over $C_{2}$, perhaps due to previously available datasets, while the evidence ratio or Bayes factor\footnote{Cosmological simulations are run one set of parameters at a time, so in the current framework we are dealing with parameter estimation as opposed to full model selection, i.e. $C_{1}$ and $C_{2}$ differ only by the value of their parameters}, $R$, is defined as:
\begin{equation}
\label{eqn:bayesfactor}
	R = \frac{P(D|C_{1})}{P(D|C_{2})}
\end{equation}
where the marginal likelihood $P(D|C_j)$ denotes the probability that one would observe data $D$ assuming a cosmology $C_j$. A large Bayes factor $R\gg1$ reflects a shift in preference for $C_{1}$ and vice-versa. Performing comparisons for many cosmological models would require numerous simulations, each run under various cosmologies; this is outside the scope of the current work. In the present work, the aim is to estimate the marginal likelihood: the probability of observing the Einstein radii of the high-z MACS sample under a single chosen hypothesis: \lcdm~with aforementioned parameters. This is non-trivial because a likelihood function related to the original observables ($\theta_{E}$ and $M_{500}$) is intractable; the finite number of objects from the simulations mean that the full $\theta_{E}$--$M_{500}$ space cannot be sampled. 

To circumvent this problem, we assume a generative probabilistic model with the form of a power-law relation between the strong lensing and mass proxies, and perform a fitting to the following function in logarithmic space\footnote{The pivot mass $9\times10^{14}\msol$ is chosen to approximate the logarithmic average of the observed and simulated clusters. Similarly the pivot Einstein radius is chosen to be 20 arcseconds.}:
\begin{equation} \label{MassERfit}
	\log \left[ \frac{M_{500}}{9\times10^{14}\msol } \right] = \alpha \log \left[ \frac{\theta_{E}}{20"} \sqrt{ \frac{D_{d}}{D_{ds}} } \right]  + \beta
\end{equation}
 with parameters ($\alpha$ and $\beta$) and aim to find the probability of observing {\it the scaling relationship}. However, rather than calculating precise values for $\alpha$ and $\beta$, one would determine a probability distribution, $P(\alpha,\beta)$, that reflects the degree of belief in their respective values. Thus $P(\alpha,\beta | D)$ acts as a summary statistic for the dataset. The relevant linear regression method, following \citet{Hogg2010}, as well as the priors on $\alpha$ and $\beta$, are outlined in Appendix~\ref{App:BayesFit}.

Next, we outline how to estimate the marginal likelihood.
In the following, ${\cal I}$ represents background information such as knowledge of the cluster selection criteria, the method of characterising the Einstein radius, and the assumption that there exists a power-law relation between strong lensing and mass.
\begin{enumerate}
\item Fit equation~(\ref{MassERfit}) to the data to obtain the posterior probability distribution, $P(\alpha, \beta | D, {\cal I})$, for $\alpha$ and $\beta$.
\item Computer simulations are run within the framework of a chosen cosmological hypothesis, $C_j$. In our case, $C_j$ represents the assumption that \lcdm~(with aforementioned values for cosmological parameters), is the true description of cosmology.
\item Simulated galaxy clusters are selected according to specified criteria, ideally reflecting the criteria used to select the real clusters. Their masses, $M_{500}$, are noted.
\item Different on-sky projections of these three-dimensional objects produce different apparent measurements of structural properties. Therefore, we construct a large number $N$ of mock samples by randomly choosing an orientation-angle and calculating $\theta_{E}$ for each cluster. The fair sampling means that for each mock dataset, denoted $D_i$ for $i$ from 1 to $N$, that $P(D_i|C_j,I) = 1/N$.
\item 
Equation~(\ref{MassERfit}) is fit to each mock sample to determine a posterior probability distribution, $P(\alpha_i, \beta_i | D_i, {\cal I})$, over $\alpha_i$ and $\beta_i$.
\item 
Consider the proposal $H_i := D_i = D$, in other words, that the mock sample resembles the real data. The resemblance is judged using the summary statistic. We assign a weight, $w_i$, to each mock sample according to its similarity to the real data:
\begin{equation}
\begin{split}
\label{eqn:weight}
w_i 	& \equiv P(H_i | D, D_i, C_j, {\cal I}) \\
	&= \int P(H_i | \alpha, \beta, D_i, {\cal I}) P(\alpha, \beta | D, {\cal I}) \diff \alpha \diff \beta \\
	&\approx \int P(\alpha_i=\alpha, \beta_i=\beta | \alpha, \beta, D_i, {\cal I}) P(\alpha, \beta | D, {\cal I}) \diff \alpha \diff \beta.
\end{split}
\end{equation}
noting that $D$ is redundant once $\alpha$ and $\beta$ are known.
The weight is equal to the integral over the product of the two (mock and real) posterior probability distributions\footnote{this is equivalent to the cross-correlation, evaluated at zero-shift}.
\item Finally, we estimate the marginal likelihood:
\begin{equation}
\begin{split}
\label{eqn:calcZ}
{\cal Z}_j 	&= \hat{P}(D|C_j)\\
		&= \sum_{i=1}^{N} P(D=D_i, D_i | C_j, I)\\
		&= \sum_{i=1}^{N} P(D=D_i | D_i, C_j, I) P(D_i | C_j, I)\\
		&\approx \frac{1}{N} \sum_{i=1}^{N} P( H_i | D, D_i, C_j, I)\\
		&= \frac{1}{N} \sum_{i=1}^{N} w_i.
\end{split}
\end{equation}
noting that $C_j$ is redundant once $D_i$ is known. Thus the marginal likelihood is estimated to be the mean weight over all mock samples.
\end{enumerate}
However, given the following equivalence:
\begin{equation}
\begin{split}
\label{eqn:calcZpractice}
\frac{1}{N} \sum_{i=1}^{N} \int P(\alpha_i=\alpha, \beta_i=\beta | D_i, {\cal I}) P(\alpha, \beta | D, {\cal I}) \diff \alpha \diff \beta \\ 
= 
 \int \Bigl[ \frac{1}{N} \sum_{i=1}^{N} P(\alpha_i=\alpha, \beta_i=\beta | D_i, {\cal I}) \Bigr] P(\alpha,\beta | D, {\cal I}) \diff \alpha \diff \beta\,
\end{split}
\end{equation}
we can add the many posteriors of all mock samples and re-normalise: $P(\alpha, \beta | C_j, {\cal I}) = \frac{1}{N} \sum_{i=1}^{N} P(\alpha_i,\beta_i | D_i, {\cal I})$, then identify ${\cal Z}$ as the zero-shift cross-correlation of the two aforementioned posteriors at the origin.
Indeed, this is our practical approach. Note that since the PDFs in equation~\ref{eqn:calcZpractice} that must be added or multiplied are initially Monte Carlo sampled, we are forced to estimate the functions on a regular 2D \{$\alpha$,$\beta$\} grid via some choice of kernel. Whether nearest grid point, gaussian convolution, or kernel density estimation, with bandwidth of $0.1<\Delta\alpha<0.5$ and $0.02<\Delta\beta<0.1$, all produce almost identical results. Any variation in ${\cal Z}$, at most 5\%, is negligible compared to the uncertainties explored in the next section.

Our solution requires simulations that produce several mock datasets, each of which can be compared to the real data via a summary statistic and kernel distance metric, as in approximate Bayesian computation \citep[ABC; see][for examples of applications of ABC within the astrophysical literature]{Cameron2012,Weyant2013,Robin2014,Ishida2015,Akeret2015,Lin2015}. It differs from standard ABC in two ways. Firstly, what we propose is not a likelihood-free approach; indeed our aim is to calculate a marginal likelihood. Secondly, rather than rejecting -- and wasting -- mock samples that are dissimilar to the real data, they are down-weighted; the weights incorporate both the kernel and the distance metric in traditional ABC. The effect should be similar to probabilistic acceptance of mock samples as outlined in \citet{Wilkinson2008} and soft ABC as mentioned in \citet{Park2015}. Model selection here would boil down to comparing mean weights rather than acceptance fraction.

While the choice of summary statistics is open to discussion (our relatively simplistic choice is justified by the small sample and large scatter), the actual metric used to define the weighting/distance is not arbitrary. The caveat is that this method is only possible in cases where one can infer the summary statistics as a probability distribution for any (mock or real) dataset.

\section{\texorpdfstring{\lcdm~strong}{LCDM strong} lensing results}
\label{Sec:Results}
In this section we calculate ${\cal Z}$, where the data $D$ are the high-z MACS clusters and the cosmological model $C$ is standard \lcdm. For an initial demonstration in Sec.~\ref{SubSec:AGNLX}, we use our fiducial simulations and methodology, while in sections \ref{SubSec:counterparts} to \ref{SubSec:relax} we examine the aspects of simulations and selection methods that affect this value. 
\subsection{\texorpdfstring{\AGNsim~clusters at $z=0.5$}{AGN clusters at z>0.5} selected by X-ray flux}
\label{SubSec:AGNLX}
The flux cut that was employed in the high-z MACS sample can be translated into a luminosity cut and using the $L_{X}$ (observer rest-frame) determined for each cluster we can select simulated clusters for our sample (see. Sec.\ref{SubSec:simclus}).
\begin{equation}
	f_{\mathrm{cut}} = \frac{L_{X,\mathrm{cut}}}{4\pi D_{L}^{2}},
\end{equation}
where $D_{L}$ denotes the luminosity distance to the cluster. For the standard \lcdm~cosmology adopted in the present work, and at $z\approx0.5$, the threshold flux translates to a threshold luminosity of $L_{X,\mathrm{cut}} =  7.6 \times 10^{44}$ erg s$^{-1}$ (observer rest-frame [0.1-2.4] keV). In the \AGNsim~simulation set, 15 clusters exceed the luminosity/flux threshold ($L_{X} > L_{X,\mathrm{cut}}$)\footnote{If more simulated clusters are available for follow-up studies, one could also account for cosmic variance}.

\begin{figure}
	\includegraphics[width=\linewidth]{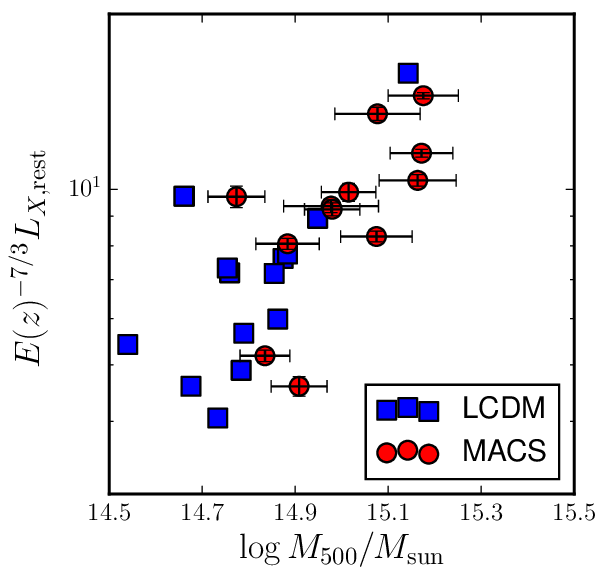}
      	\caption{Cluster rest-frame X-ray luminosity, computed within the [0.1-2.4] keV energy band, as a function of cluster mass $M_{500}$. Blue squares denote $z=0.5$ clusters selected from the \AGNsim~simulations, while red circles denotes the high-z MACS sample.
	}
	\label{LXMz050}
\end{figure}

There exists a tight correlation between core-excised X-ray luminosities of the high-z MACS clusters and X-ray based mass-estimates \citep{Mantz2010}. We do not use core-excised luminosities to select our simulated clusters, however, nor do we use bolometric luminosities in the cluster rest-frame, but rather the [0.1-2.4] keV band luminosities in the observer's frame. This is done to best replicate the actual selection criteria.
In Fig.~\ref{LXMz050} we show the \LMrel~relation for the simulated and observed clusters, where the X-ray luminosities are measured in the soft X-ray waveband. The \LMrel~self-similar relation is technically correct for bolometric luminosities, {\it but} \citet{RB02} also suggest a relation with L$_{X}$ in our band \citep{Perrenod80}.  Self-similarity is generally better followed by relaxed clusters in hydrostatic equilibrium, while a high-luminosity sample would be biased towards unrelaxed clusters; for such a sample, one might expect higher luminosities for a fixed mass.
Even with the inclusion of AGN feedback, the simulated clusters remain slightly overluminous for a given mass, relative to the observed clusters. There is a distinct lack of high mass candidates in the simulated sample, witnessing that for the adopted cosmology even a box size as large as 1 h$^{-1}$Gpc does not contain a large enough population of massive clusters at the redshift of interest, $z=0.5$. Note that in the following analysis, we do not simply consider the observed distribution of Einstein radii, which would be severely biased by the lack of high-mass clusters, but rather the lensing-mass relation.

\begin{figure*}
	\label{fig:fiducial}
	\includegraphics[width=\linewidth]{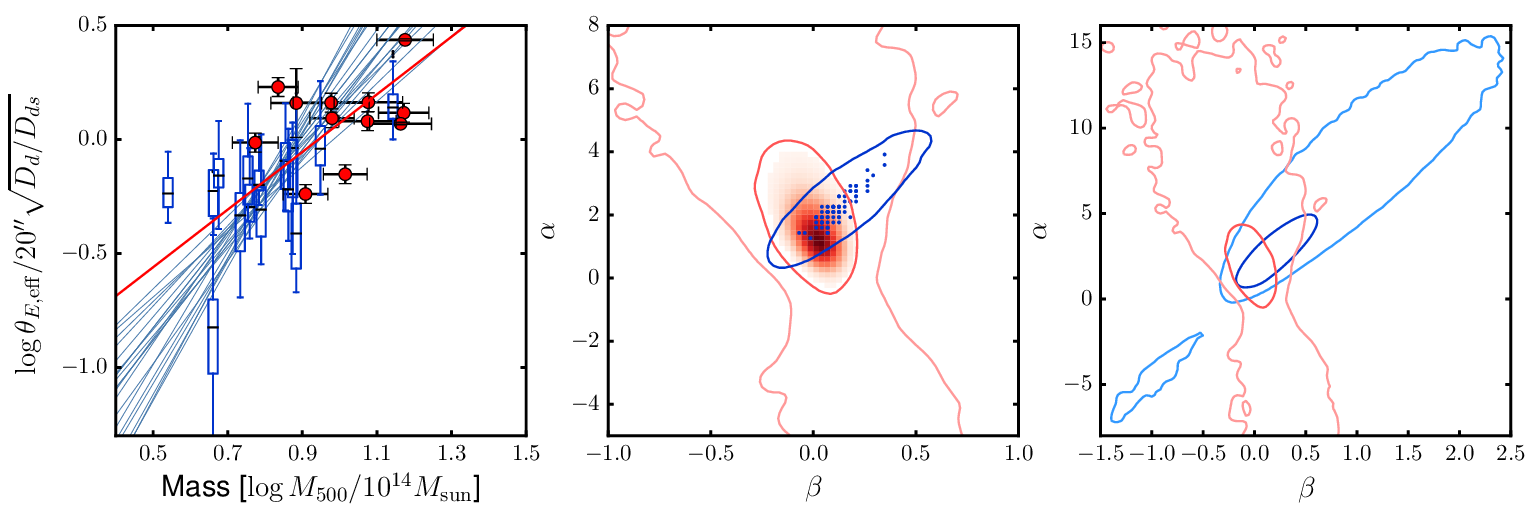}
      	\caption{Einstein radii statistics for $z=0.5$ clusters from the \AGNsim~simulations. 
	{\it Left}: Strong lensing efficiency, characterised by scaled Einstein radii, $\EinRadEff$, plotted as a function of $M_{500}$. The range of Einstein radii for simulated clusters are shown by the blue box-plots. The red circles represent the MACS $z>0.5$ clusters, whose masses have been corrected as described in Sec.~\ref{Sec:MACS}. The red line marks the maximum a-posteriori fit to observational data, while the thin blue lines mark the fit to 20 randomly chosen mock samples from simulations. 
          {\it Middle}: 1-$\sigma$ and  2-$\sigma$ constraints on parameters of the strong lensing -  mass relation given the MACS $z>0.5$ cluster data (red contours and shading). Overplotted in blue dots are the best fits to 80 mock observations of $z=0.5$ clusters from the \AGNsim~simulations. A typical 1-$\sigma$ error is shown as a blue ellipse. 
	{\it Right}: Same as the middle panel, but the blue curves mark the 1-$\sigma$ and 2-$\sigma$ contours of $P(\alpha, \beta | C, {\cal I})$, having combined all mock observations. The value of ${\cal Z}$} is the integral over the product of the PDFs marked by the red and blue contours.
\end{figure*}
%
In the left panel of Fig.~\ref{fig:fiducial} we show the relation between the Einstein radii and the cluster mass $M_{500}$. The $z>0.5$ clusters of the MACS sample are represented by red circles. For simulated clusters, the situation is more complicated. Since different lines of sight provide a large variation in projected mass distribution, each cluster cannot be associated with an individual Einstein radius, nor a simple Gaussian or log-normal distribution \citep[see][]{Killedar12}. We therefore measure the Einstein radius for 80 different lines of sight and, for ease of visualisation, describe the distribution of Einstein radii for each simulated cluster by a box-plot\footnote{In the box-plots, we mark the median with a short black horizontal line, a blue box marking the 25th and 75th percentiles and stems to meet the furthest data-points within 1.5  times the inter-quartile range}.

As described at the beginning of this section, we fit the observational data to the strong lensing-mass relation and after marginalising out the nuisance parameter, $V$, present $P(\alpha, \beta | D, {\cal I})$, the posterior distribution for $\alpha$ and $\beta$, denoted by red contours in the middle panel of Fig.~\ref{fig:fiducial}. 
 Many mock samples are individually fit to the lensing-mass relations; the maximum of the posterior is shown as a blue point, one for each mock, and a typical 1-$\sigma$ error shown as a blue ellipse. By adding the posteriors for each mock sample and renormalising, we estimate $P(\alpha, \beta | C, {\cal I})$, shown by the blue contours in the right-hand panel of Fig.~\ref{fig:fiducial}. By multiplying by the two distributions, we find ${\cal Z} = 0.25$.

Note that one cannot comment on whether the marginal likelihood is {\it large} or {\it small}. One cannot use this value to claim `consistency' or `tension' with \lcdm. However, if the same process is repeated for simulations under a different cosmological model then the Bayes factor $R$ can be calculated and, after accounting for priors, it may (or may not) reveal a preference for one of the cosmologies, in light of this data. 

Note that we have allowed for negative $\alpha$, i.e. negative slopes for the strong lensing-mass relation, which seems counterintuitive but should not be ruled out on principle given the anti-correlation between concentration and cluster mass. If we do insist on positive slopes, by placing a boundary on the prior on $\alpha$, we find that ${\cal Z}$ increases by about 5\%. 

The use of the \AGNsim~simulation set and the cluster selection as described characterise our fiducial approach. Certain factors can be expected to impact the marginal likelihood, and consequently the Bayes factor when comparing cosmologies. For the remainder of this section, we consider how the marginal likelihood for our specific cosmology may depend on other details, such as cluster redshift, selection criteria and the numerical implementation of baryonic processes.

\subsection{Effect of baryon processes}\label{SubSec:counterparts}
By using hydrodynamic simulations, we are now in the position of being able to select clusters in a manner more consistent with the selection of the observational sample. However, given the sub-grid nature of the astrophysical processes and subsequent uncertainties in their implementation, we consider it prudent to determine the sensitivity of our scientific conclusions to the inclusion of baryonic processes and resulting gas distribution (see \citealt{Wurster13} and \citealt{RagoneFigueroa13} for detailed discussions on different implementations of AGN feedback). In this section we perform the same comparison as before but for the clusters in the \DMsim, \NRsim~and \CSFsim~simulations. The clusters selected are the counterparts to the clusters selected in the previous subsection Sec.~\ref{SubSec:AGNLX}. That is, clusters are selected in the \AGNsim~simulation according to their X-ray luminosity as would be observed in the 0.1--2.4 keV band; then, the {\it same} clusters are selected in the other simulations\footnote{Note that these clusters are not selected based on their X-ray luminosity in the \DMsim, \NRsim~and \CSFsim~simulations; the degree to which $L_{X}$ is overestimated in \NRsim~and \CSFsim~simulations is greater than for \AGNsim~simulations, and X-ray luminosities are obviously not defined for the \DMsim~clusters.}. Therefore, we are able to determine the effect of baryonic physics on the marginal likelihood estimated using the same sample of clusters.

\begin{figure*}
\begin{minipage}{0.325\textwidth}
	\includegraphics[width=\linewidth]{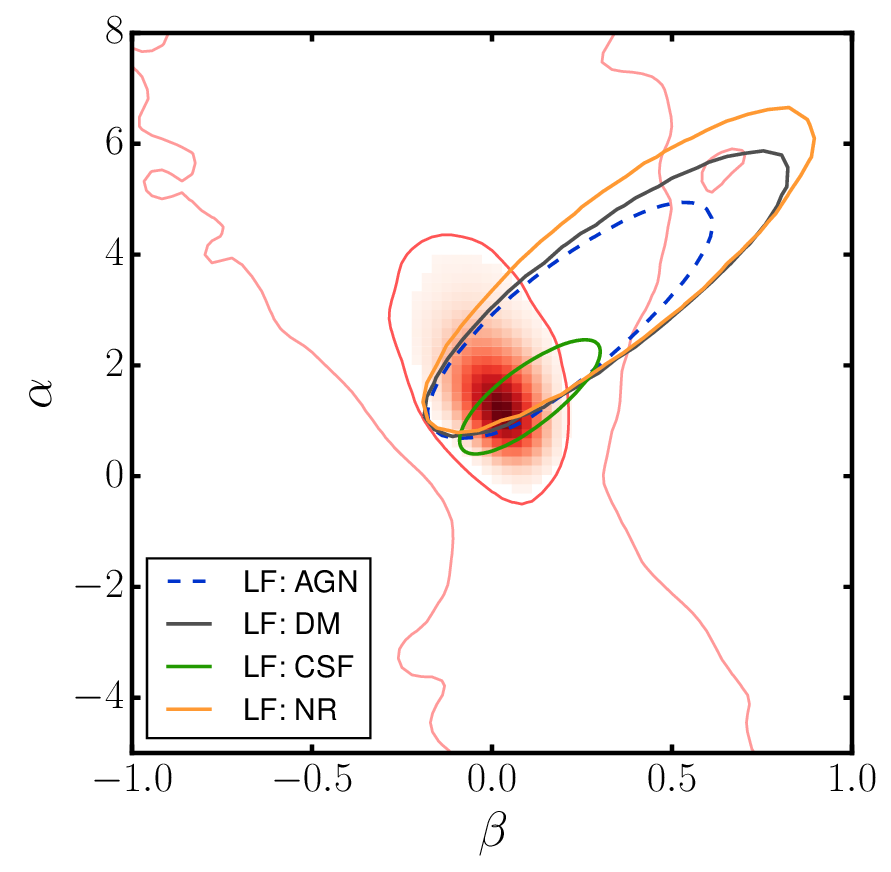}
\end{minipage}
\begin{minipage}{0.325\textwidth}
	\includegraphics[width=\linewidth]{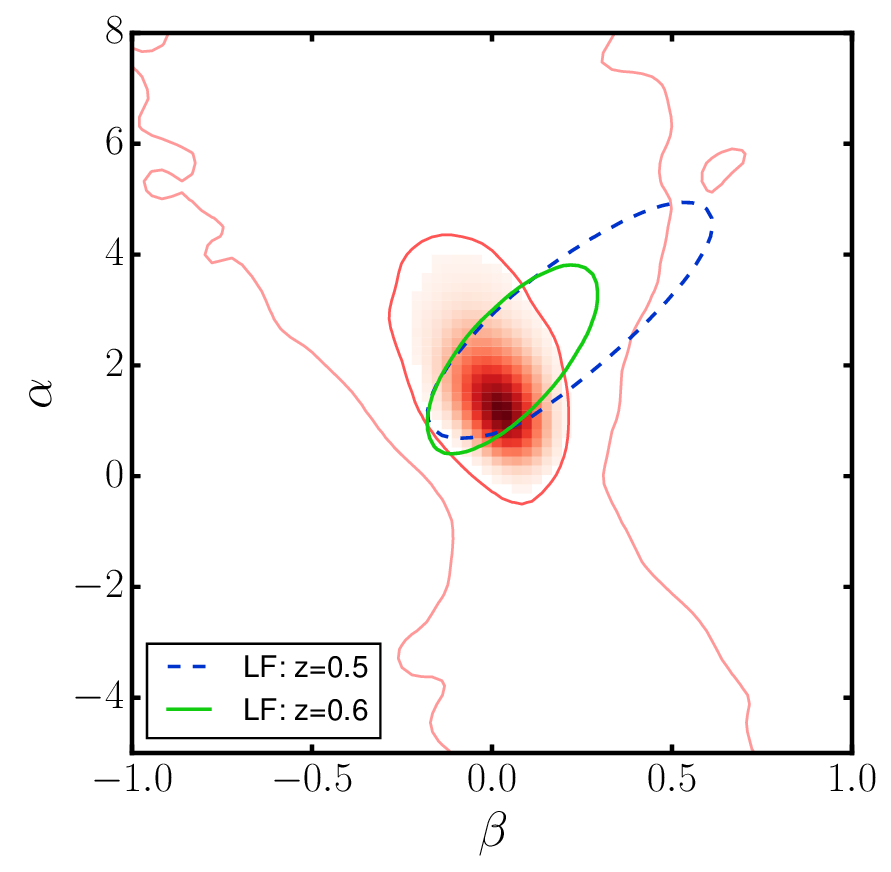}
\end{minipage}
\begin{minipage}{0.325\textwidth}
	\includegraphics[width=\linewidth]{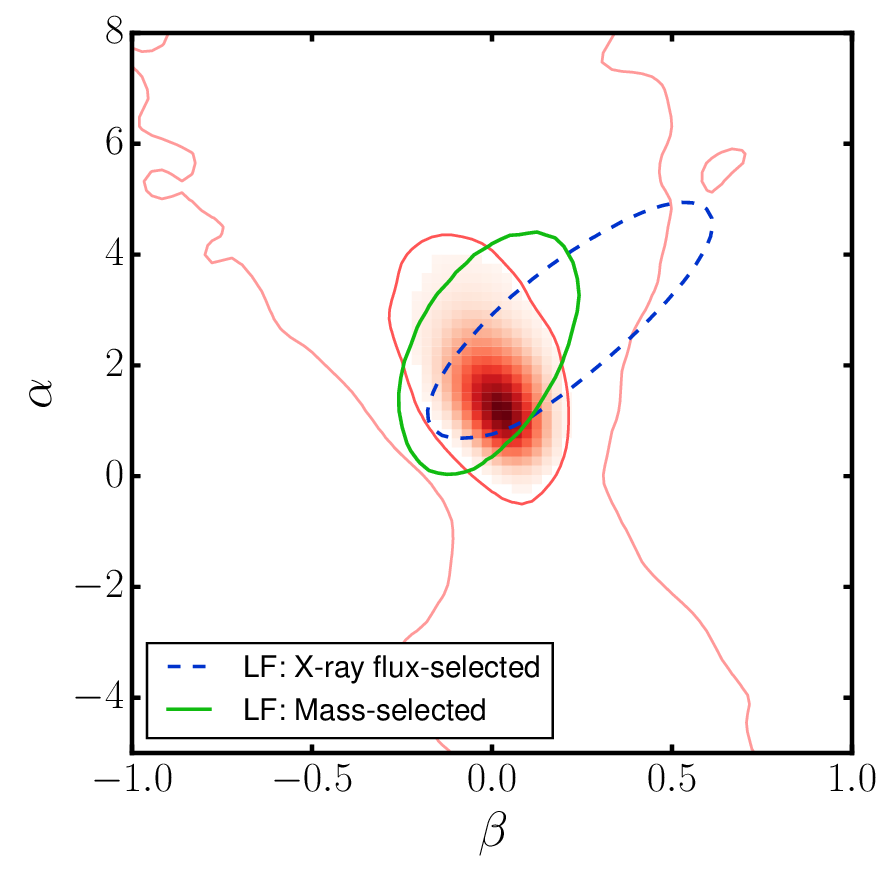}
\end{minipage}
      	\caption{Observed fit to the scaling relation as in the right-hand panel of Fig.~\ref{fig:fiducial} (red) and 1-$\sigma$ contours of $P(\alpha, \beta | C, {\cal I})$ derived from: 
({\it Left panel}) the different simulation sets: {\AGNsim}~(blue; fiducial), {\DMsim}~(grey), {\CSFsim}~(green), {\NRsim}~(orange);
({\it Middle panel}) simulated clusters at: $z=0.5$ (blue; fiducial) and $z=0.6$ (green);
({\it Right panel}) simulated clusters selected by:  X-ray flux (blue; fiducial) and mass (green) 
}
	\label{fig:ab_Alternatives}
\end{figure*}

Radiative processes can impact on cluster mass, $M_{500}$, as well as the mass profiles and consequently, strong lensing at $z=0.5$ \citep{Killedar12,Cui2014}. \CSFsim-like simulations, which suffer from overcooling and steepened profiles, result in up to 10\% increase in $M_{500}$ relative to dark matter simulations at $z=0.5$, with no significant trend with mass, but increase the Einstein radii {\it for low-mass haloes in particular}, resulting in a preference for a lower value of $\alpha$, as seen in the left-hand panel of Fig.~\ref{fig:ab_Alternatives}. Halos are more spherical as a consequence of adiabatic contraction; the increased degree of sphericity causes, in turn, a decrease in the variation between mock samples, and therefore a much narrower $P(\alpha, \beta | C, {\cal I})$ distribution. AGN feedback tempers the effect on both cluster mass and strong lensing. Ultimately, we find ${\cal Z} = 0.74$ for the \CSFsim~simulations, about triple that derived from the \AGNsim~simulations, while the \DMsim~and \NRsim~simulations result in ${\cal Z} = 0.18$ and 0.14 respectively, about half that of \AGNsim\footnote{As an aside, we could assume that the cosmological model is correct and test cluster physics instead using such simulations but in that case perhaps the posterior for $\alpha$ and $V$ would be more revealing.}.  These simulations are extremes in terms of the astrophysical processes that are ignored. However, as they bracket our true ignorance of the thermal and kinetic effects of baryonic processes, then this implies an uncertainty in the ${\cal Z}$ -- and therefore $R$ -- of a factor of three.

\subsection{Changing the redshift of simulated clusters}\label{SubSec:hiz}
Eleven out of the twelve high-z MACS clusters lie within $0.5<z<0.6$ (J0744.8+3927 lies at $z\approx0.7$). Thus far we have compared the strong lensing properties of the MACS clusters with simulated clusters by extracting the latter at the lower end of the redshift range: $z=0.5$.  We remind the reader that the Einstein radius is scaled in order to account for the different redshifts of the clusters being
compared. However, this does not account for any structural differences due to clusters at $z=0.5$ being captured at a later stage of evolution than those at $z=0.6$. Since cluster-mass-concentration at fixed total mass is expected to increase with redshift, then the choice of simulating clusters at only $z=0.5$ could potentially underestimate the \lcdm~prediction for strong lensing for clusters where this redshift is only the lower-limit. Therefore, we repeat the comparison for simulated clusters at $z\approx0.6$ instead. At $z=0.6$, the threshold flux translates to a threshold luminosity of $L_{X,\mathrm{cut}} = 11.9 \times 10^{44}$ erg s$^{-1}$ (observer rest-frame [0.1-2.4] keV); eleven clusters in the \AGNsim~simulations satisfy this selection criteria.

The simulation-based $P(\alpha, \beta | C, {\cal I})$ and the observed $\alpha$ and $\beta$ are shown in the middle panel of Fig.~\ref{fig:ab_Alternatives}, akin to the right-hand panel of Fig.~\ref{fig:fiducial}. 
The relationship between the Einstein radii and mass is similar to that of the $z=0.5$ simulated sample, notwithstanding the absence of any cluster with unusually small Einstein radius. The result is a more strongly peaked function $P(\alpha, \beta | C, {\cal I})$ (blue contours). In this case, we measure ${\cal Z} = 0.49$, which is almost double that derived from the $z=0.5$ simulations.

\subsection{Cluster selection by mass}\label{SubSec:mass}
It is common practice to select simulated cluster samples using a mass threshold, or some other proxy for X-ray flux, when simulations do not contain gas dynamics \citep[e.g.][]{H10, Men11}. However, since the clusters in the MACS survey (among others) were selected by flux rather than luminosity, there is no corresponding mass threshold, strictly speaking. In fact, even if the selection is by luminosity, there will be a preference for high-concentration clusters, for a fixed mass, which are relatively X-ray brighter \citep{Rasia13}.

On the other hand, mimicking X-ray selection for simulated clusters requires a robust treatment of the hot X-ray emitting ICM. While our simulated clusters follow the luminosity-temperature relation reasonably well \citep{Planelles14}, we note that there is still some small degree of over-luminosity in the scaling relation against mass, possibly due to violation of hydrostatic equilibrium in observational mass estimates. The single-redshift choice means that flux, luminosity and mass thresholds are equivalent. Thus we are able to investigate if the selection of clusters by X-ray luminosities introduces low-mass clusters into our simulated sample. To address this concern, we select simulated clusters by imposing a $M_{500}$ mass threshold corresponding to the lowest mass $z>0.5$ MACS cluster.

Assuming that there are not many low-luminosity high-mass clusters introduced into the sample, we could expect, given the left panel of Fig.~\ref{fig:fiducial}, that in fact a few MACS clusters would have Einstein radii that are smaller than those typically measured in simulated clusters. Accounting for the 2009 {\it Chandra} calibration update, we estimate the lowest mass high-$z$ MACS cluster to have $M_{500}=4.2\times10^{14} h^{-1}\msol$. Accordingly, we select the eight clusters from the {\it AGN} simulations with $M_{500}$ above this value. Indeed, we find that compared to the simulated sample from Sec.~\ref{SubSec:AGNLX}, the seven lowest mass clusters drop out of the simulated sample, while no low-luminosity clusters are added. The function $P(\alpha, \beta | C, {\cal I})$ for both assumed selection methods and observational fit to the scaling relation are shown in the right-hand panel of Fig.~\ref{fig:ab_Alternatives}. By selecting simulated clusters by mass, ${\cal Z}$ increases to 0.42.

\subsection{Cluster selection by dynamical state}\label{SubSec:relax}
Five of the twelve $z>0.5$ MACS clusters are classified as dynamically relaxed, according to a morphological code described in \citet{E07}.
This low fraction is not surprising for an X-ray flux-selected sample since cluster mergers are known to lead to large boosts in X-ray luminosity \citep[e.g.][]{RT02,Planelles2009}. On the other hand, observational programmes such as CLASH collect clusters according to their X-ray contours and alignment of the BCG with the X-ray peak, in an effort to choose relaxed clusters. Here, we determine how this choice of relaxed clusters affects the marginal likelihood.
 
\begin{figure*}
	\includegraphics[width=\linewidth]{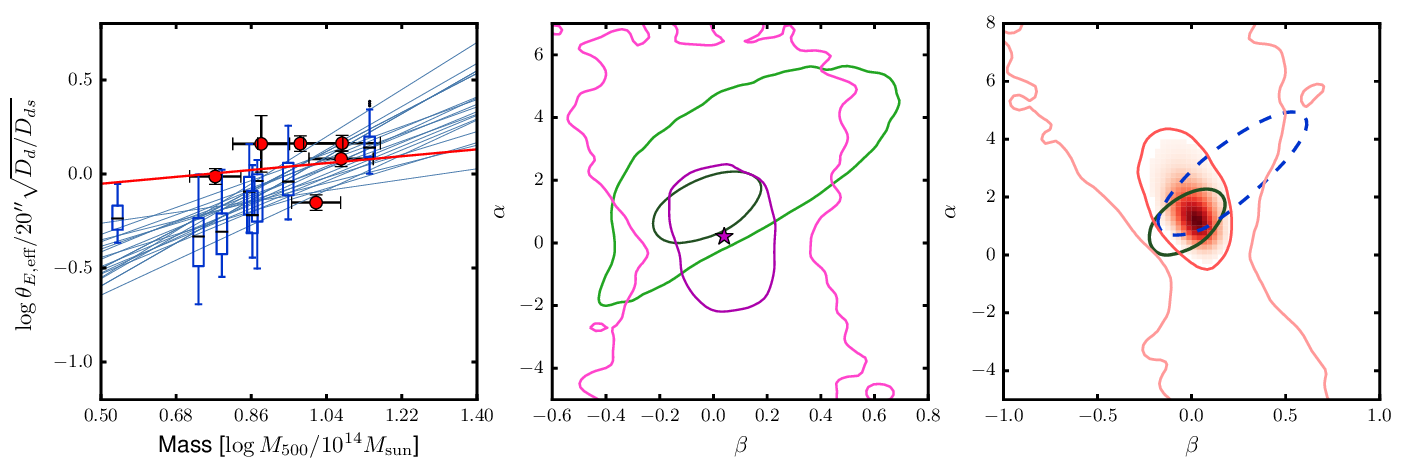}
      	\caption{ {\it Left}: As in the left-hand panel of Fig.~\ref{fig:fiducial} but for dynamically relaxed clusters;
	{\it Middle}: Constraints on $\alpha$ and $\beta$ for the relaxed sub-sample of the MACS clusters (magenta; best-fit marked with a star), and $P(\alpha, \beta | C, {\cal I})$ using the relaxed simulated clusters (green);
          {\it Right}: Constraints on $\alpha$ and $\beta$ for the complete MACS sample (red), and $P(\alpha, \beta | C, {\cal I})$ using the complete sample of simulated clusters (blue dashed) and the relaxed sub-sample (green solid)}
	\label{fig:relaxed}
\end{figure*}

The background information $\cal I$ now includes the dynamical state criterion, and so the mock data are limited to the relaxed sub-sample out of the fiducial \AGNsim~clusters.
The dynamical state of the simulated clusters is defined following the method of \citet{Killedar12}. The method consists of computing the offset between the position of the particle with the minimum gravitational potential and centre of mass (COM), where the COM is calculated within a range of radii $\zeta_i \rvir$, with $\zeta_i$ going from $0.05$ to $2$ in 30 logarithmic steps. A cluster is defined as relaxed if the offset is less than 10 per cent of $\zeta_i \rvir$ for all radii.  In the \AGNsim~simulations 9 relaxed clusters exceed the luminosity/flux threshold ($L_{X} > L_{X,\mathrm{cut}}$). In both the MACS and simulated sample, half the clusters are deemed relaxed. The Einstein radii and masses of the relaxed sub-sample of clusters are shown in the left-hand panel of Fig.~\ref{fig:relaxed}; with the observed fit described by the magenta contours and $P(\alpha, \beta | C, {\cal I})$ shown in green in the middle panel. The inferred marginal likelihood is ${\cal Z} = 0.39$.
%

It would be unwise to use theoretical models based on the assumption of a relaxed sample {\it if} the observational sample did not include this criterion. In order to demonstrate this, consider now the effect of applying the incorrect selection criteria when modelling lenses. Excluding the most disturbed clusters from both the simulated and observational sample might tighten the fit on $\alpha$ and $\beta$, but this is compensated for by a loosening fit due to the smaller sample size. Ultimately, as seen in the right-hand panel of Fig.~\ref{fig:relaxed}, the function $P(\alpha, \beta | C, {\cal I})$ derived from the relaxed simulated clusters (green contours) is more sharply peaked than that which is derived without this additional selection (blue contours). If this relaxed simulated sample is used to analyse the full observational sample (constraints shown in red), one would incorrectly derive a value of ${\cal Z} = 0.7$; we remind the reader that including merging clusters in the simulated sample resulted in a smaller marginal likelihood by a factor of four.

Relaxed sub-samples as shown here provide much less data to work with, and thus a less powerful cosmological test. Furthermore, the dynamical state of simulated clusters has been evaluated in three dimensions, which is not exactly consistent with choices based on projected observables. Criteria presented by \citet{Merten2015}, \citet{Meneghetti2014} and the automated methods of \citet{Mantz2015} have recently made it possible to mimic the morphological selection. If this can be applied to a large number of both simulated and observed clusters with mass estimates and strong lensing measurements, we could determine a more robust marginal likelihood.


\section{Comparison to previous work}\label{Sec:CompareLit}
There have been four other works in the literature that analyse the strong lensing statistics of the high-z MACS clusters and compare them to predictions from simulations \citep{Horesh11,Men11,Z11a,Waizmann2014}. Several factors could lead to disagreements with our findings presented in Sec.~\ref{Sec:Results}. Firstly, the theoretical predictions for the first three works were based on the adoption of 1-year {\it Wilkinson Microwave Anisotropy Probe} \citep[WMAP-1][]{Spergel2003} cosmological parameters including, most notably, a high normalisation for the matter power spectrum: $\sig=0.9$. These parameters predict an earlier epoch for structure formation relative to the preferred model based on WMAP-7 results \citep{Komatsu2011}, and thus clusters are predicted to be more concentrated and stronger lenses than predicted by our simulations. 
Additionally, simulations in the above previous analyses were either collisionless N-body, or included only non-radiative physics; however as we have shown in Sec.~\ref{SubSec:counterparts}, the effects of baryons are minor compared to the substantial scatter associated with cluster triaxiality.
Finally, the inferred strong lensing properties of the high-z MACS sample for all three studies described below were based on mass models constructed prior to the availability of the high quality HST data within the CLASH and Frontier Fields programmes.
All these factors aside, our main focus here is on the statistical methods used.

\citet{Horesh11} measured the frequency of arc production in cluster lenses from the {\tt Millenium} simulation \citep{Springel2005} at three different redshift bins simulated clusters ($z\sim$0.2, 0.4 and 0.6). The high-z MACS sample were compared with the highest redshift bin. They found that simulations under-predicted the number of arcs per cluster, but cautioned that there were too few simulated clusters available (4 and 1 for their low and high mass threshold respectively) to form a robust conclusion.

\begin{figure}
	\includegraphics[width=\linewidth]{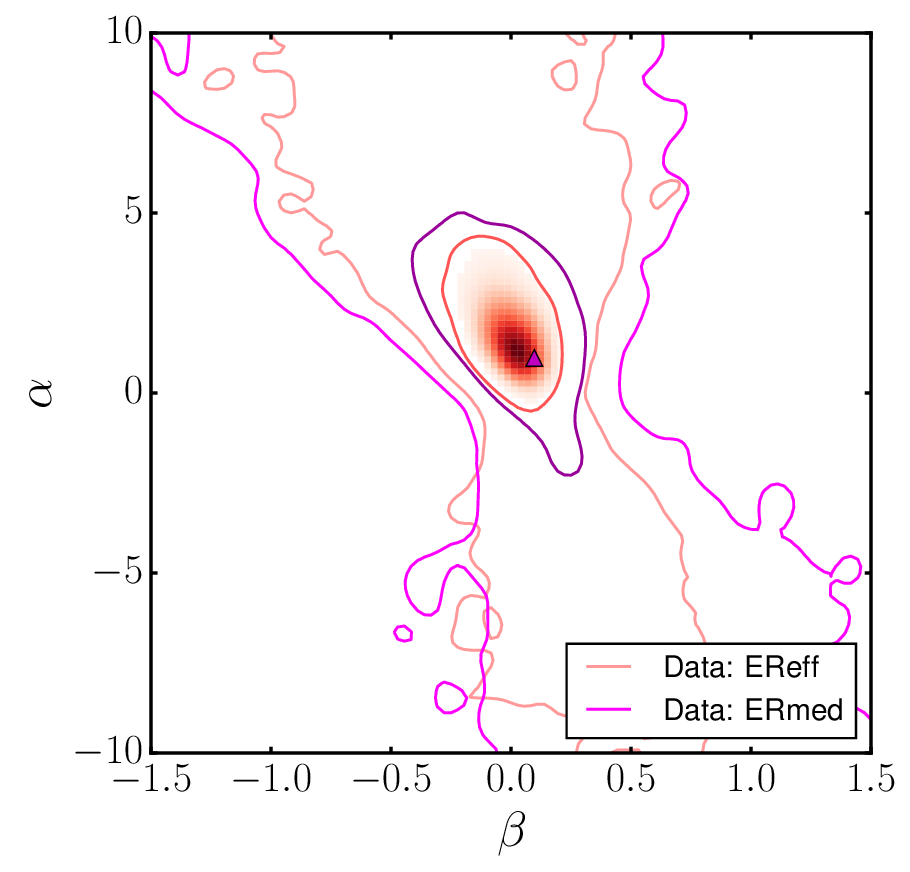}
      	\caption{The posterior probability on $\alpha$ and $\beta$ for the high-z MACS data for Einstein radii defined as $\EinRadEff$ (red; best fit marked with a star) and $\EinRadMed$ (magenta; best fit marked with a triangle)}
	\label{fig:ab_ERmed}
\end{figure}

\citet{Men11} based the \lcdm~predictions on clusters from six snapshots between $0.5<z<0.7$ from the {\tt MareNostrum Universe} non-radiative gas simulations \citep{Gottlober2007}. The selection criteria included X-ray flux selection, but required a correction term for luminosities that are estimated from simulations that have a relatively simple description of the gas. Their comparisons revealed that the predicted lensing cross-section for giant arc-like images of sources at $z_{s}=2$ was half that of the observed value\footnote{The values of $\EinRadMed$ as quoted in~\citet{Men11} are based on the earlier mass models, prior to the availability  of HST data within the CLASH and Frontier Fields programmes.}, while the Einstein radii (characterised through the alternative `median' radius) differed by 25 per cent. They then claimed to close the gap between observations and simulations with the inclusion of realistically modelled merging clusters. However, \citet{Redlich12} have shown that the median Einstein radius is more sensitive to cluster mergers and will be boosted for a longer period during the merger. In fact, three of the MACS clusters were singled out for their unusually strong lensing qualities: MACS J0717.5+3745, MACS J0025.4-1222 and MACS J2129.4-0741; yet these are precisely the clusters for which the characterisation of the Einstein radii makes the most difference: $\EinRadMed/\EinRadEff$ = 1.3, 1.9 and 2.2 respectively.
Fig.~\ref{fig:ab_ERmed} demonstrates that there is a much poorer scaling relationship between the Einstein radius if characterised in this alternative manner and the cluster mass for the $z>0.5$ MACS clusters due to merger-driven boosts.
%

\citet{Z11a} determined a theoretical distribution for effective Einstein radii using smooth triaxial lens models with parameters constrained by N-body simulations. The lens structure was designed to match results from earlier collisionless simulations and they selected clusters above a mass threshold based on the flux-limit using the (relaxed cluster) $L_{X}$--$M$ relation from \citet{RB99}. Einstein radii were found to be about 1.4 times larger than predicted by \lcdm, measured by comparing the medians of the total distributions.

A recent study by \citet{Waizmann2014} takes a different approach to the strong lensing comparison by using order statistics. Semi-analytic models of cluster lensing and mass function are combined with general extreme value distributions to determine `exclusion constraints' on the $n$-largest observed Einstein radii; the high-z MACS sample is found to be consistent with \lcdm.

The studies of \citet{Horesh11} and \citet{Men11} employ directly the results from N-body and non-radiative gas simulations, while \citet{Z11a} and \citet{Waizmann2014} use semi-analytic models. In the present work, we have modelled cluster lenses from hydrodynamic simulations, allowing a realistic description of baryonic effects and of unrelaxed clusters, as well as the directly comparable X-ray flux selection. However, unlike the semi-analytic models, the number of objects is still limited.

The key difference between the present study and those conducted before lies in the statistical approach. \citet{Waizmann2014} employs order-statistics, while the first three studies employ the Kolmogorov-Smirnov (KS) test in order to compare the distribution of a strong lensing property (either lensing cross-sections or Einstein radii) for the high-z MACS sample to the similarly constructed distribution for a mock sample consisting of simulated clusters; otherwise the median of their distributions are compared as a consistency check. However:
\begin{enumerate}
\item Comparing medians of distributions involves the loss of much of the information in the data. 
\item Uncertainties in the measurement of the Einstein radii are ignored.
\item Different lines of sight through the same simulated cluster can produce different mock observations due to cluster triaxiality and the presence of substructures, leading to wildly varying results for the KS-test
\item All aforementioned statistical approaches --- and the majority of the literature analysing the strong lensing efficiency or concentration of galaxy clusters --- do not formally provide a marginal likelihood that would allow one to judge the preference for the \lcdm~cosmological model  {\it over other cosmologies}.
\end{enumerate}

In the present work instead, we take a Bayesian approach and provide a guide to performing the first step of the model-selection problem: determining the marginal likelihood, the probability of observing the lensing-mass relationship assuming a single power-law form to the scaling relation. The rest of the comparison requires numerous simulations for other cosmologies and is therefore outside the scope of this work. However, we have laid the groundwork for a strong lensing test of cosmology.


\section{Triaxial models}\label{Sec:triax}

Beta-profile fits to X-ray emission from the ICM provide a gas-mass estimate which is converted to a total mass estimate assuming a gas/baryon fraction in clusters; the mass recovered is limited to within $R_{500}$. Weak lensing data can allow one to measure the total mass of a cluster out to larger radii and perhaps include shape measurements in combination with X-ray and/or Sunyaev-Zel'dovich (SZ) data \citep{Marshall2003,Mahdavi2007,Morandi10,Sereno2011,Sereno2013,Limousin13}. Triaxial model fits provide more realistic descriptions of cluster mass profiles, but they will generally also result in a higher mass estimate than if spherical symmetry was assumed \citep{CK07}. Given the increasing availability of lensing and SZ data, and subsequent possibility of shape reconstruction, we consider it useful to explore the relationship between strong lensing efficiencies of clusters and their ellipsoidal mass. We have performed triaxial shape measurements on the simulated clusters for a fixed overdensity of 200 and 2500 to derive axis ratios (see Appendix \ref{App:Triax} for details) and thus calculate the resulting mass within the ellipsoid (M$_{200,\mathrm{tri}}$ and M$_{2500,\mathrm{tri}}$ respectively).

\begin{figure*}
	\includegraphics[width=\linewidth]{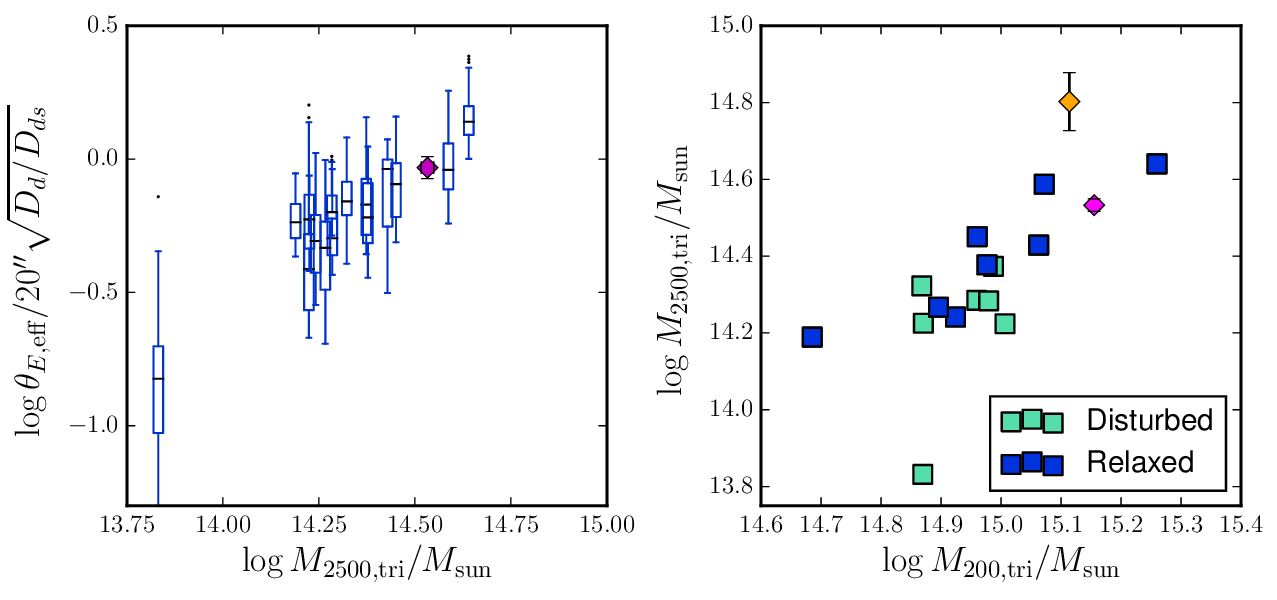}
      	\caption{Einstein radius as a function of enclosed mass within ellipsoid for an average overdensity of $\Delta=2500$. Blue box plots represent simulated clusters, while the magenta diamond represents MACS J1423.8+2404, and the orange diamond Abell 1689}
	\label{triM}
\end{figure*}

Despite the large line-of-sight related variation, the Einstein radius of a cluster tends to scale well with mass at high overdensities as seen clearly in the left-hand panel of Fig.~\ref{triM}. Therefore, we may use the mass at high-overdensities in place of Einstein radius for a scaling relation as previously discussed. Consider, then the potential to use the relationship between triaxial masses at low and high overdensities, as shown in the right-hand panel of Fig.~\ref{triM}, and the associated fit parameters as summary statistics. The differences between mock samples is no longer due to triaxiality, but rather due to cosmic variance. A full simulation-based likelihood function will require a much more extensive cluster sample, so unfortunately it is out of the scope of the present work.

As for observational data, we consider the following two galaxy clusters. MACS J1423.8+2404 is arguably the most dynamically relaxed cluster of the high-z MACS sample and has a relatively low-substructure fraction \citep{Limousin10}. This has made it an ideal candidate for reconstruction of the triaxial mass profile using multiple data-sets. \citet{Morandi10} have reconstructed the 3-dimensional structure of this cluster using a combination of X-ray, SZ and weak lensing data, determining a best-fit set of triaxial parameters and the enclosed mass for an overdensity of $\Delta=2500$. \citet{Limousin13} improved the algorithm used to combine data and provide a best fit set of generalised NFW model parameters and enclosed mass for an overdensity of $\Delta=200$. A second cluster, Abell 1689 is not part of the high-z MACS sample but has been the subject of extensive study regarding its triaxial shape and mass reconstruction using a variety of probes \citep{Sereno2011,Sereno2013,Limousin13,Umetsu2015}. We show both these clusters in Fig.~\ref{triM} as well, although there are too few to draw any strong conclusions. At present, multi-wavelength reconstruction methods are likely to be restricted to relaxed clusters, so we would advocate providing likelihood functions for relaxed clusters (as marked by dark blue squares) as well.


\section{Conclusions}\label{Sec:Conclude}

The MACS $z>0.5$ sample has been the subject of a number of studies in the literature \citep[i.e.][]{Horesh11,Men11,Z11a,SZ12} regarding the consistency of their strong lensing properties with \lcdm. Since these studies were undertaken, several of the clusters have had revisions to their mass models as a result of the multi-band data and precise HST imaging from the CLASH and Frontier Fields programmes; in some cases the new Einstein radii are significantly lower than previously estimated. 

Our primary goal here is to step away from claims of `tension' or `consistency' with a single cosmological model. We have introduced a new approach to calculating a marginal likelihood for measuring the relationship between strong lensing efficiency and cluster lens mass, and demonstrated with the high-z MACS clusters under the \lcdm~model and a choice of cosmological parameters consistent with WMAP-7 results. Model predictions are based on a finite sample of galaxy clusters simulated with hydrodynamic processes. Since our interest lies in typical strong lensing properties of cluster lenses at fixed mass, we consider a scaling relation between the Einstein radius of the lens and its mass. We assume a power-law relation (see equation~\ref{MassERfit}), and interpret the parameters of the scaling relations, $\alpha$ and $\beta$, as summary statistics. We employ cosmological simulations to determine the probability of observing these parameters under \lcdm. Folding this function with the observed fit, we derive the marginal likelihood. This forms the first step in a Bayesian model selection process.

The approach described above is an exciting new strategy for estimating the marginal likelihood for a given cosmology using strong lensing galaxy clusters. {\it However}, we recognise that the calculation involves running computer simulations that can take months. We find that the marginal likelihood would vary by a factor of three or so only if the most unrealistic simulations are employed, with significant overcooling or otherwise lacking radiative processes entirely. We invite development of cosmological simulations that include a large range of hydrodynamical processes with different philosophies for the implementation of sub-grid physics, feedback models and implementation of hydrodynamical schemes.  We expect simulations run under a range of cosmological models to be analysed in a manner equivalent to that demonstrated in the present work to allow eventual model selection. Our findings suggest that if a model-comparison study was carried out using a simulation based on an alternative cosmological hypothesis and resulted in a Bayes factor of 20 or more (see equation~\ref{eqn:bayesfactor}), then \DMsim~simulations would be sufficient. However, in the event that the $R$ is found to be smaller, then the computationally expensive hydrosimulations would be necessary.

Cluster selection is an important factor, however, particularly the choice by dynamical state. Samples that include clusters in unrelaxed dynamical states are problematic: they introduce a large amount of scatter in the lensing-mass scaling relation, which consequently weakens the constraints on $\alpha$ and $\beta$ and reduces the power of cosmological test. On the other hand, relaxed sub-samples are smaller in number, which have a similar effect. Furthermore, since some measures of strong lensing (e.g. $\EinRadMed$) can be highly sensitive to merging events (and line of sight on the sky), smaller samples will suffer more from occasional boosts in lensing efficiency, and weaken the strength of statistical tests. We warn against using theoretical models that assume a relaxed morphology to analyse an observational sample that includes clusters undergoing mergers. Instead, we recommend the use of effective Einstein radius to characterise strong lensing in a large homogeneous sample of relaxed clusters selected with automated methods, such as those of \citet{Mantz2015}. 

The relationship between masses at low and high overdensities is an interesting alternative to the strong lensing-mass scaling relation. Here, we advocate folding cluster triaxiality into the analysis in order to reduce variation between mock samples. Measuring these masses will become feasible over the coming years given the availability of multi-wavelength data. Alternatively, weak lensing estimates alone would provide a more precise measure of cluster mass, albeit usually at lower overdensities or at fixed aperture radius \citep[e.g.][]{Applegate2014}.

For follow-up studies, we aim to:
\begin{enumerate}
\item include more simulated clusters in a parent sample that would then allow one to account for `cosmic variance' between mock samples, not just the orientation-related scatter. This would reduce any bias in the estimator of the marginal likelihood.
\item use semi-analytic models or emulators in place of individual simulated clusters \citep[e.g.][]{Giocoli2012,Kwan2013,Bonamigo2015,Jabot2014}. This would allow even easier modelling of cosmic variance, however, fitting-functions here are still based on simulations.
\item relax the assumption of the power-law relationship between the Einstein radius and cluster lens mass. However a new summary statistic would need to be identified, possibly with more than two dimensions. Otherwise, alternative definitions of discrepancy distance could be explored \citep[e.g.][]{Park2015}.
\item apply generalised linear models \citep[GLMs, e.g.][]{deSouza2015a,Elliott2015,deSouza2015b} in place of the analysis as described in Sec.~\ref{Sec:ML}, in the case that full error distributions on measurements become available.
\end{enumerate}

While numerical simulations have opened up the possibility to explore the effects of cosmology and physics on non-linear structure formation, they have brought with them the necessity to break away from `textbook' statistical methods that assume the existence of analytical models. We hope, therefore, that we have begun a discussion about alternatives which may prove as revolutionary to cosmological testing as the simulations themselves.

\section*{Acknowledgements}
The authors thank Johannes Buchner, Ewan Cameron, Geraint Lewis, Jiayi Liu, Hareth Saad Mahdi, Julio Navarro, Dennis Prangle, Matthias Redlich, Christian Robert, Alexandro Saro, Micha Schirmer, Mauro Sereno, Anja von der Linden and Adi Zitrin for helpful discussions, as well as Giuseppe Murante for his involvement in running the simulations. We also thank the anonymous referee for comments that aided in clarifying the manuscript.
MK acknowledges support by the DFG project DO 1310/4-1, and a fellowship from the European Commission's Framework Programme 7, through the Marie Curie Initial Training Network CosmoComp (PITN-GA-2009-238356). K.D. acknowledges the support by the DFG Cluster of Excellence ``Origin and Structure of the Universe". SP acknowledges support by a grant funded by the ``Consorzio per la Fisica di Trieste", by the PRIN-INAF09 project ``Towards an Italian Network for Computational Cosmology'', by the PRIN-MIUR 201278X4FL grant ``Evolution of Cosmic Baryons'', by the INDARK INFN grant, and by the Spanish Ministerio de Econom\'ia y Competitividad (MINECO; grant AYA2013-48226-C-2-P) and Generalitat Valenciana (grant PROMETEO II/2014/069).
Simulations have been carried out at the CINECA supercomputing Centre in Bologna (Italy), with CPU time assigned through ISCRA proposals and through an agreement with University of Trieste. This work has been supported by the PRIN-MIUR 2012 grant ``Evolution of Cosmic Baryons'', by the PRIN-INAF 2012 grant ``Multi-scale Simulations of Cosmic Structure" and by the INFN ``INDARK'' grant. We acknowledge support from "Consorzio per la Fisica - Trieste". 
This work has made use of the following open-source software: python\footnote{http://www.python.org}, numpy\footnote{http://www.numpy.org}, scipy\footnote{http://www.scipy.org}, matplotlib \citep{Hunter2007} and cosmolopy\footnote{http://roban.github.com/CosmoloPy/}. The bibliographic research was possible thanks to the tools offered by the NASA Astrophysical Data Systems.


\bibliographystyle{mnras} 
\bibliography{Killedar_PML}


\appendix

\section{Fitting to the lensing-mass scaling relation}\label{App:BayesFit}
We employ a Bayesian fitting procedure to determine the summary statistics, $\alpha$ and $\beta$, parameters of the scaling relation between strong lensing efficiency and total cluster mass (equation~\ref{MassERfit}). We also acknowledge that there is likely to be intrinsic scatter in this relationship directly comparable to the scatter in the concentration-mass relation, partly due to cluster triaxiality and substructure and partly from the varying formation histories of the clusters. Thus we also include a nuisance parameter, $V$, which represents intrinsic Gaussian variance, orthogonal to the line.

For this appendix, we change notation in order to reduce the subscripts: the mass of the $i$-th cluster lens as $M_i$, and the scaled Einstein radius --- however characterised --- as $E_i$. Each
data-point is denoted by the vector $\bmath{Z}_i = [ \log M_i ,
\log E_i ]$. Their respective uncertainties ({\it on the
  logarithms}) are denoted $\sigma_{M}^{2}$ and $
\sigma_{E}^{2}$. Since we assume the uncertainties for Einstein radii
and cluster mass are uncorrelated, the covariance matrix,
$\bmath{S}_i$, reduces to:
\begin{equation}
 \bmath{S}_i \equiv 
 \left( \begin{array}{cc}
 \sigma_{M}^{2} & 0 \\
 0 & \sigma_{E}^{2}
 \end{array} \right)
\end{equation}
In the case of a mock sample of simulated clusters, $\bmath{S}_i=0$. We expect uncertainties in mass due to the choice of halofinder and Einstein radii (see our method describe in sec.~\ref{SubSec:ERcalc}) to be negligible compared to the overall scatter.

Consider now the following quantities: $\varphi \equiv \arctan \alpha$, which denotes the angle between the line and the x-axis, and $b_{\perp} \equiv \beta \cos \varphi$ which is the orthogonal distance of the line to the origin. The orthogonal distance of each data-point to the line is:
\begin{equation}
\Delta_i = \hat{\bmath{v}}^{\top}   \bmath{Z}_i  - \beta \cos \varphi
\end{equation}
where $\hat{\bmath{v}} = [ -\sin\varphi, \cos\varphi ] $ is a vector orthogonal to the line.

Therefore, the orthogonal variance is 
\begin{equation}
    \Sigma_i^{2} = \hat{\bmath{v}}^{\top}   \bmath{S}_i  \hat{\bmath{v}}\,.
\end{equation}

Following \citet{Hogg2010} (see their Eqn.~35), we calculate the (logarithm of the) likelihood over the 3-dimensional parameter space $\bmath{\Theta_{1}} \equiv \{\alpha, \beta, V \}$ given the data $D$, which includes $\bmath{Z}$ and $\bmath{S}$:
\begin{equation}\label{likeXerrYerr}
    \ln {\cal L}(\bmath{\Theta_{1}} ; D) = \mathrm{K} - \sum_{i=1}^{N}\frac{1}{2}\ln(\Sigma_i^{2} + V)  - \sum_{i=1}^{N}\frac{\Delta_i^{2}}{2\Sigma_i^{2} + V}
\end{equation}
where K is an arbitrary constant, and the summation is over all clusters
in the considered sample.

While we ultimately (aim to) provide the parameter
constraints on $\alpha$ and $\beta$, flat priors for these tend to
unfairly favour large slopes. A more sensible choice is flat for the
alternative parameters $\varphi$ and $b_{\perp}$. We apply a modified
Jeffreys prior on $V$:
\begin{equation}
\label{eqn:priorV}
\pi(V) \propto \frac{1}{V+ V_{t}}
\end{equation}
This is linearly uniform on $V$ for small values and logarithmically uniform on $V$ for larger values with a turnover, $V_{t}$, chosen to reflect the typical uncertainties.

Thus, for each $\bmath{\Theta_{1}}$, we may define an alternative set of parameters $\bmath{\Theta_{2}} \equiv \{ \varphi , b_{\perp} , V \}$, for which the prior is given by:
\begin{equation}
\begin{split}
     \pi(\bmath{\Theta_{2}}) 	&=   \pi(\varphi , b_{\perp})\pi(V) \\	
     					&\propto \pi(V)
\end{split}
\end{equation}
where $\pi(V)$ is given by equation~(\ref{eqn:priorV}).
The prior on $\bmath{\Theta_{1}}$ is then dependent on the magnitude of the Jacobian of the mapping between the two sets of parameters:
\begin{equation}
\begin{split}
     \pi(\bmath{\Theta_{1}}) & =   \pi(\bmath{\Theta_{2}})\mathrm{det} \frac{\partial \bmath{\Theta_{2}}}{\partial \bmath{\Theta_{1}}} \\
     			 	 	&\equiv
  \pi(\bmath{\Theta_{2}})\frac{1}{(1+\alpha^{2})^{3/2}}
\end{split}
\end{equation}
Boundaries on the priors are sufficiently large\footnote{The physically motivated choice of restricting $\alpha\geq0$ is also explored, however this has very minor effects on the final results despite removing the (small) secondary peak in the marginal posterior on $\alpha$ and $\beta$}: $-8\leq\beta\leq8$; $-40\leq\alpha\leq40$; $0\leq V\leq V_{\mathrm{max}}$. $V_{\mathrm{max}}$ is chosen to reflect the overall scatter in the data.
The posterior is calculated following Bayes' theorem:
\begin{equation}
     P(\bmath{\Theta_{1}} | D)  \propto  {\cal L}(\bmath{\Theta_{1}} ; D) \,  \pi(\bmath{\Theta_{1}})
\end{equation}
and is normalised. Since we are interested in the constraints on $\alpha$ and $\beta$, we then marginalise over the nuisance parameter, $V$.

In practice, the posterior distribution was sampled by employing {\tt emcee} \citep{ForemanMackey2013}, the python implementation of the affine-invariant ensemble sampler for Markov chain Monte Carlo (MCMC) proposed by \citet{GW2010}. The proposal distribution took the form of a small Gaussian `ball' centred on the expected peak in the distribution as determined in earlier test runs. These tests also found that the auto-correlation times tend to be between 10 and 60 steps for each parameter. For both the data and each of the 80 mock samples, we ran 120 walkers for 600 steps each. In all cases, the 120 initial steps were considered to encompass the `burn-in' phase and were discarded.

\section{Triaxiality and enclosed mass}\label{App:Triax}

Given a numerically simulated dark matter halo for which the underlying mass distribution is described by the distribution of discrete particles, its shape can be determined with the following iterative procedure, based on \citet{Dubinski1991} (see also a detailed discussion in \citealt{Zemp2011}): 
 \begin{enumerate} 
 \item compute and diagonalise the inertia shape tensor 
 \begin{equation}
 S_{ij} \equiv \frac{\Sigma_{k} m_{k} w_{k} x_{k,i} x_{k,j}}{\Sigma_{k} m_{k}}
 \end{equation}
 where $m_{k}$ is the mass of the $k$th particle, $w_{k}$ is a weighting associated with that particle and $x_{k,i}$ is the $i$ component of the particle position vector. The summation is over all particles chosen for analysis.
 \item identify the principal axes as the eigenvectors of $S_{ij}$.
 \item calculate the eigenvalues of $S_{ij}$ ($\lambda_{1}>\lambda_{2}>\lambda_{3}$).
 \item determine the axis ratios using
 \begin{equation}\label{eigenvsaxisratios}
	\frac{\lambda_{3}}{\lambda_{1}} =  s^{2}
	\qquad \mathrm{and} \qquad 
 	\frac{\lambda_{2}}{\lambda_{1}} = q^{2} .
 \end{equation}
 \item identify a new set of particles for analysis by calculating $r_{e}$ for each particle based on the newly determined $q$ and $s$. If $r_{e}$ is less than a pre-determined $r_{e}^{\star}$, the particle is included in the integration volume.
 \item repeat steps i--iv until both $q$ and $s$ pass convergence
   criteria or until a maximum number of iterations.
 \end{enumerate}
 
The choice of $r_{e}^{\star}$ determines the scale of the ellipsoidal volume at each iteration. We perform the iterative procedure outlined above with $r_{e}^{\star}$ defined such that the enclosed mass has a fixed average density of $\Delta$ times the critical density. Note that numerous works in the literature scale the ellipsoid at each iteration by anchoring the major axis to a fixed radius instead, but while our approach is more numerically demanding, this criteria better reflects the required information about cluster shape \citep[see also][]{Despali13, Bonamigo2015}. 
 We repeat this process two times for each cluster, once for $\Delta=200$ and then $\Delta=2500$.

\bsp
\label{lastpage}
\end{document}